\documentclass[aps,twocolumn,nofootinbib,amsmath,amssymb]{revtex4}

\usepackage[utf8]{inputenc}

\usepackage{graphicx}
\usepackage{subfig}

\usepackage{natbib}

\usepackage{bm}

\usepackage{url}
\usepackage{hyperref}
\hypersetup{colorlinks=true, citecolor=black, filecolor=black, linkcolor=black, urlcolor=black}

\usepackage{siunitx} 

\begin{document}

\title{A Wide Dataset of Ear Shapes and Pinna-Related Transfer Functions Generated by Random Ear Drawings}


\author{\href{https://orcid.org/0000-0002-9823-335X}{\includegraphics[scale=0.06]{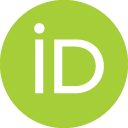}\hspace{1mm}Corentin \surname{Guezenoc}}}
\email{corentin.guezenoc@centralesupelec.fr}
\affiliation{FAST Research Team \\IETR (CNRS UMR 6164) \\CentraleSup\'elec \\Avenue de la Boulaie, 35510 Cesson-S\'evign\'e, France}
\author{\href{https://orcid.org/0000-0001-7199-7563}{\includegraphics[scale=0.06]{orcid.png}\hspace{1mm}Renaud \surname{S\'eguier}}}
\email{renaud.seguier@centralesupelec.fr}
\affiliation{FAST Research Team \\IETR (CNRS UMR 6164) \\CentraleSup\'elec \\Avenue de la Boulaie, 35510 Cesson-S\'evign\'e, France}



\begin{abstract}
Head-related transfer functions (HRTFs) individualization is a key matter in binaural synthesis.
However, currently available databases are limited in size compared to the high dimensionality of the data. 
Hereby, we present the process of generating a synthetic dataset of 1000 ear shapes and matching sets of pinna-related transfer functions (PRTFs), named WiDESPREaD (wide dataset of ear shapes and pinna-related transfer functions obtained by random ear drawings) and made freely available to other researchers. 
Contributions in this article are three-fold. 
First, from a proprietary dataset of 119 three-dimensional left-ear scans, we build a matching dataset of PRTFs by performing fast-multipole boundary element method (FM-BEM) calculations. 
Second, we investigate the underlying geometry of each type of high-dimensional data using principal component analysis (PCA). 
We find that this linear machine learning technique performs better at modeling and reducing data dimensionality on ear shapes than on matching PRTF sets. 
Third, based on these findings, we devise a method to generate an arbitrarily large synthetic database of PRTF sets that relies on the random drawing of ear shapes and subsequent FM-BEM computations. 
\end{abstract}

\maketitle

\section{Introduction}
In daily life we unconsciously capture the spatial characteristics of the acoustic scene around us thanks to auditory cues such as sound level, time-of-arrival and spectrum. 
Such cues derive from the alterations of sound on its acoustic path to our eardrums, which depend not only on the room and the position of the acoustic source, but also on the listener's morphology.
Their mathematical description in free-field is called head-related transfer functions (HRTFs) in the frequency domain and head-related impulse responses (HRIRs) in the time domain \citep{moller_fundamentals_1992}.
They are the cornerstone of a technique called binaural synthesis that allows the creation of a virtual auditory environment through headphones: by convolving a given sound sample with the right pair of HRIRs before presenting it to the listener, the sound sample is perceived at the desired location.

The use of a non-individual HRTF set in binaural synthesis is known to cause discrepancies such as wrong perception of elevation, weak externalization and front-back inversions \citep{wenzel_localization_1993}.
Thus, a lot of work has been done for the past decades towards user-friendly HRTF individualization,
among which four categories can be identified \citep{guezenoc_hrtf_2018}. 
Acoustical measurement \citep{wightman_headphone_1989-1} is the state-of-the-art method and relies on a heavy measurement apparatus and is time-intensive.
Numerical simulation allows the simulation of HRTFs from a 3D scan of a listener's morphology \citep{kahana_boundary_1998}. 
The associated individual measurement phase is much less troublesome in terms of equipment (a portable light-based scanner can be used for instance),
nevertheless the approach is time-intensive, particularly so during the simulation step. 
Finally, the two latter families of approaches aim at providing somewhat low-cost but real-time solutions to the matter. 
They are usually based either on anthropometric measurements \citep{middlebrooks_individual_1999, zotkin_customizable_2002, hu_head_2006, hu_hrtf_2008}
or on perceptual feed-back \citep{middlebrooks_psychophysical_2000, seeber_subjective_2003, hwang_modeling_2008-1, yamamoto_fully_2017}
and often rely heavily on HRTF databases.

However, currently available measured HRTF databases \citep{algazi_cipic_2001, majdak_3-d_2010, watanabe_dataset_2014, carpentier_measurement_2014, bomhardt_high-resolution_2016, brinkmann_cross-evaluated_2019} are small compared to the dimensionality of the data.
Indeed, the largest that we know of, the ARI (acoustics research institute) database \citep{majdak_3-d_2010}, features 120 subjects, while the dimension of a typical high-resolution HRIR set \citep{bomhardt_high-resolution_2016} is about $1.2 \cdot 10^6$ (256 time-domain samples $\times$ 2300 directions $\times$ 2 ears). 
While work has been done towards combining existing databases \citep{andreopoulou_towards_2011, tsui_head-related_2018},
such composite databases can hardly attain the same level of homogeneity as a database made in a single campaign. 
Furthermore, the total number of subjects would amount to a few hundreds at best.
Synthetic datasets have also been built by numerically simulating HRTF sets from scans of listener morphology \citep{rui_calculation_2013, jin_creating_2014, kaneko_ear_2016, brinkmann_cross-evaluated_2019}.
However, to the best of our knowledge, only the HUTUBS database \citep{brinkmann_cross-evaluated_2019} is fully public.
Moreover, such datasets are not larger than acoustically measured ones.
Indeed, the largest that we know of, HUTUBS \citep{brinkmann_cross-evaluated_2019}, features 96 subjects which is less than the ARI one.
This can be explained by the fact that, although less tedious than acoustic measurements, the acquisition of morphological scans for a large number of human subjects is far from trivial.

In this paper, we aim at alleviating the lack of large-scale datasets.
First, in Section~\ref{sec:numericalSimulations}, we supplement a dataset of 119 3D human left-ear scans with the corresponding 119 simulated PRTF sets.
Then, in Sections~\ref{sec:pcaEars}, \ref{sec:pcaPrtfs} and \ref{sec:compPcaModels}, we investigate the underlying geometry and the potential for dimensionality reduction of both types of data by performing principal component analysis (PCA) on each dataset.
Although it is a coarse machine learning technique whose limitations include linearity, PCA\footnote{PCA is closely related to the Karhunen-Lo\`eve transform, widely used in the field of information theory.} is a good starting point thanks to its algorithmic simplicity and high interpretability.
Finally, based on our findings, we present in Section~\ref{sec:databaseGeneration} a method to generate an arbitrarily large synthetic database of PRTF sets, which relies on random ear shape drawings and numerical acoustic simulations.

Let us note that, while we focus here on ear shapes and matching PRTFs,
the information contained in PRTFs is key to the matter of HRTF individualization.
Indeed, pinnae have a vast influence on the spectral features involved in perceptual discrepancies due to a lack of individualization \citep{asano_role_1990}.
Furthermore, pinnae constitute the most complex component of HRTF-impacting morphology, in terms of shape, inter-individual variability and in terms of how small physical changes can have a strong influence on the resulting filters.

\section{Original Ear Shape Dataset}  \label{sec:originalEarShapeDataset}
Work presented in this article is based on a proprietary dataset of left ear 3-D scans of ${n} = 119$ human subjects.
The dataset was constituted in previous work by Ghorbal, S\'eguier and Bonjour \citep{ghorbal_method_2019}.
The pinna meshes were acquired using a commercial structured-light-based scanner.
They were then normalized in size and rigidly aligned.
Finally, they were registered: the point clouds were re-sampled so that they were in semantic correspondence with each other, 
thus sharing an identical number of vertices ${n_v} = 17176$.
35750 triangular faces were defined based on the indices of the ${n_v}$ vertices: the definition of the faces is identical from one mesh to the other.

In the following, we denote by $E = \left\lbrace \mathbf{e}_1, \; \dots \; \mathbf{e}_n \right\rbrace$ the set of $n$ ear point clouds whose $x$, $y$ and $z$ coordinates were concatenated into row vectors  $\mathbf{e}_1, \; \dots \; \mathbf{e}_n \in \mathbb{R}^{3 {n_v}}$, with $3 {n_v} = 54528$.
As mentioned above, the only change from one mesh to the other resides in the coordinates of the ${n_v}$ vertices.
Therefore, the term `ear shape' is hereon meant as ear point cloud.

\section{Numerical Simulations of PRTFs} \label{sec:numericalSimulations}
For all ear shape $\mathbf{e}_i$ in $E$, we simulated numerically the corresponding PRTF set $\mathbf{p}_i \in \mathbb{C}^{{n_f} \times {n_d}}$,  where ${n_f}$ and ${n_d}$ denote respectively the number of frequency bins and the number of directions of measurements. 
Simulations were carried out using the fast-multipole boundary element method (FM-BEM) \citep{gumerov_fast_2005}, thanks to the mesh2hrtf
software developed by the ARI team \citep{ziegelwanger_mesh2hrtf:_2015, ziegelwanger_numerical_2015}.

We denote 
$\varphi: \mathbb{R}^{3 {n_v}} \longrightarrow  \mathbb{C}^{{n_f} \times {n_d}}$ 
the process of going from a registered ${n_v}$-vertex ear point cloud to the corresponding simulated PRTF set, which is described in the rest of the subsection.

Simulations were made for ${n_f} = 160$ frequencies from $0.1$ to 16~kHz, regularly spaced with a step of 100~Hz.
The frequency resolution was chosen so that it was finer than the equivalent rectangular bandwidth (ERB)-based frequency scale in most of the frequency range.
Indeed, the ERB scale is appropriate for HRTFs according to \citep{breebaart_perceptual_2001} and the 100-Hz-spaced linear scale is finer than the ERB scale for frequencies above 700~Hz, which is more than sufficient in the case of PRTFs, who include little spectral variations below 4-5~kHz.

\subsection{Mesh closing and grading}
First, we derived the ear mesh from the ear point cloud by incorporating the 35750 triangular faces defined by the indices of the $n_v$ vertice, as explained in Section~\ref{sec:originalEarShapeDataset}.

Second, we closed the ear mesh by 
filling the canal hole based on our prior knowledge of the boundary's vertex indices,
and then by stitching the resulting mesh onto a cylindrical base mesh.
Using such a small base mesh instead of one of a head and torso has consequences: spectral features that are usually found in HRTFs are altered (head shadowing effect is reduced to a smaller angular zone and shifted to higher frequencies) or absent (ripples due to the torso). 
However, as we did not have at our disposal a dataset of individual 3-D head and torso scans, in the latter case we would only have been able to use a generic head and torso mesh, which would have mixed non-individual spectral features with the individual pinna-related ones, at the cost of a great increase in required computing resources.
These steps were scripted in Blender\footnote{\url{https://www.blender.org/}}  Python and performed automatically using various Blender built-in mesh treatments.
 
Third, a re-sampling (also called grading) of the mesh was performed.
This step is a pre-requirement to any boundary element simulation:
the mesh ought to be as regular as possible and sampled finely enough with regard to the maximum simulated frequency.
According to Gumerov, O’Donovan, Duraiswami and Zotkin \cite{gumerov_computation_2010}, for instance, the mesh should present a uniform vertex distribution, equilateral triangles and at least five elements per wavelength.
In our case, we used the progressive grading approach proposed by
Ziegelwanger, Kreuzer and Majdak in \citep{ziegelwanger_priori_2016} and made available on-line as an OpenFlipper\footnote{\url{http://www.openflipper.org/}} plug-in,
which makes the mesh finer near the ear canal (where the sound source is positioned) and progressively coarser elsewhere.
This considerably decreases the computing cost of the FM-BEM simulation compared to uniform re-sampling, while maintaining numerical accuracy.
%

\begin{figure} 
	\includegraphics[width=0.5\textwidth]{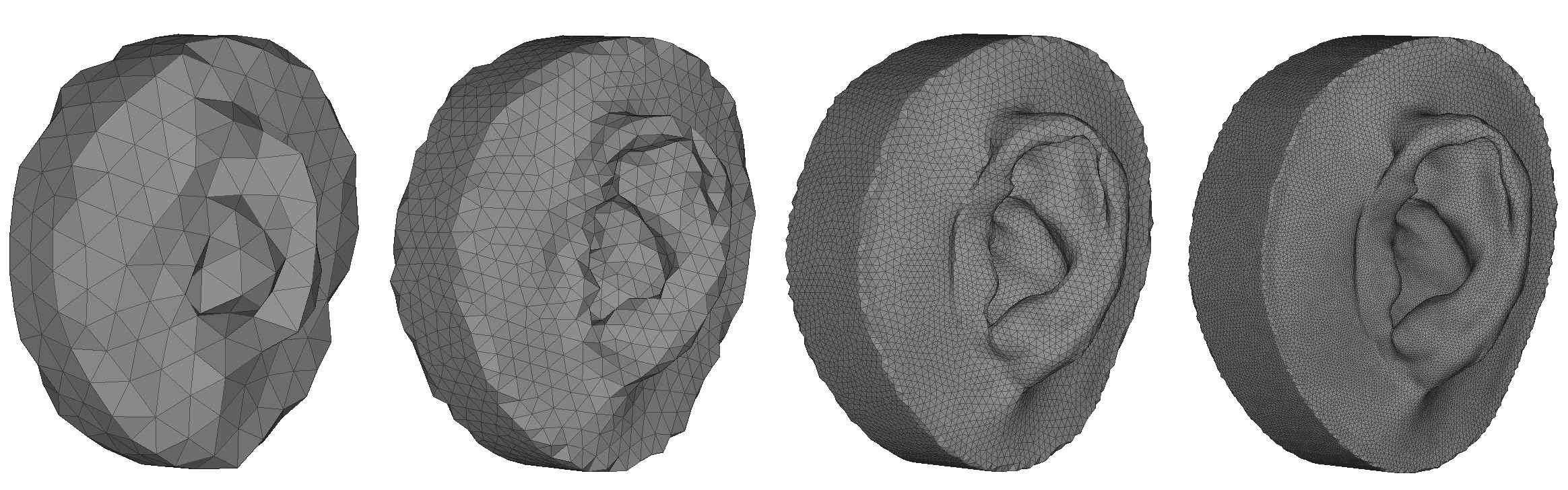}
	\caption{Simulation-ready meshes derived from ear point cloud $\mathbf{e}_1$ for four mesh grading configurations, each corresponding to a frequency band.
	Left to right: [0.1, 0.4~kHz], [0.5, 2.0~kHz], [2.1, 3.5~kHz] and [3.6, 16~kHz].}
	\label{fig:1}
\end{figure}

Additionally, in order to further reduce the computational cost, we adapted the mesh grading step to each of four different frequency bands.
At low frequencies, a uniform re-sampling was enough due to the low number of required elements.
It was performed with target edge lengths of 10 and 5 mm, in the frequency bands [0.1, 0.4~kHz] and [0.5, 2.0~kHz], respectively.
At higher frequencies, the re-sampling was progressive, with target minimum and maximum edge lengths of 2 and 5~mm, and 0.7 and 5~mm, in the frequency bands [2.1, 3.5~kHz] and [3.6, 16~kHz], respectively.
An example of simulation-ready meshes (each corresponding to a mesh grading configuration) is displayed in Figure~\ref{fig:1}.

%

\subsection{Simulation settings}
\begin{figure} 
	\includegraphics[width=0.27\textwidth]{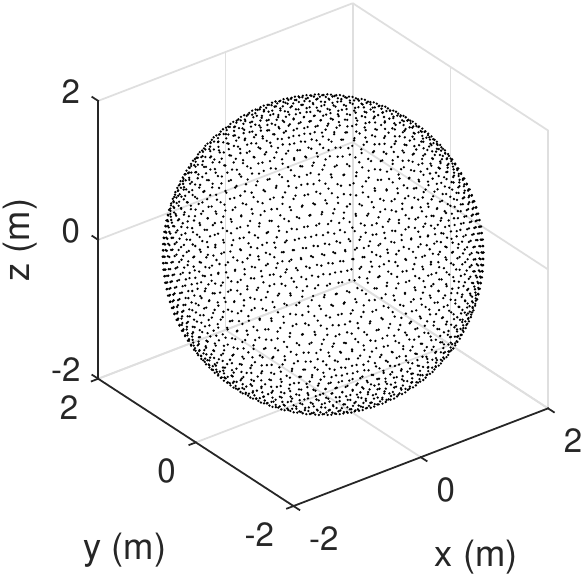}
	\caption{Spherical grid used for PRTF simulations: 2-meter-radius icosahedral geodesic polyhedron of frequency 256 (${n_d} = 2562$ vertices).}
	\label{fig:2}
\end{figure}

According to the reciprocity principle \citep{zotkin_fast_2006}, a sound source was placed at the entry of the filled ear canal, while virtual microphones were disposed on a spherical grid centered on the pinna:
a 2-meter-radius icosahedral geodesic polyhedron of frequency 256 (${n_d} = 2562$ directions), displayed in Figure~\ref{fig:2}.
Although not relevant to the rest of this article, PRTFs were computed for other virtual microphone grids as well, which are included in WiDESPREaD.


The sound source was created by assigning a vibrant boundary condition to a small patch of triangular faces located on the ear canal plug.
Elsewhere on the mesh, the boundary condition was set to infinitely reflective.
This boundary condition setting is used in the vast majority of work involving HRTF calculations \citep{kreuzer_fast_2009, jin_creating_2014}, which makes our work more easily comparable to the literature.
Furthermore, although modeling the acoustic impedance of the skin as infinite may be somewhat unrealistic, to the best of our knowledge no alternative has been proposed so far, possibly because of the limited frequency range of impedance tube measurements: up to 6.4~kHz for a standard device and up to 12.8~kHz for the experimental one proposed by Kimura, Kunio, Schumacher and Ryu \citep{kimura_new_2014}.

\subsection{Post-processing} \label{subsec:postProcessing}
Once the simulation of a PRTF set was complete, it was post-processed as follows.
Let $\mathbf{p}_{\mathrm{raw}} \in \mathbb{C}^{{n_f} \times {n_d}}$ be a PRTF set simulated for ${n_f} = 160$ frequency bins that exclude the constant component and for $n_d$ vertices of a spherical grid.

First, PRTFs were padded in frequency zero with the 100-Hz complex values.

Then, diffuse field equalization \citep{middlebrooks_individual_1999} was performed by removing the non-directional component, called Common Transfer Function (CTF), from the PRTFs. 
For all frequency bin $f = 1, \; \dots \;  {n_f}$ and for all direction of index $d = 1, \; \dots \; n_d$, 
\begin{equation}
	\mathbf{p} (f, d) = \frac{ \mathbf{p}_{\mathrm{raw}} (f, d) }{ \mathbf{c} (f) }      , \quad
\end{equation}
where the CTF $\mathbf{c} \in \mathbb{C}^{n_f}$ was obtained by calculating a Voronoi-diagram-based \citep{augenbaum_construction_1985} weighted average of the log-magnitude spectra of $\mathbf{p}$ over all directions $d = 1, \; \dots \; n_d$, then by deriving the corresponding minimal phase spectrum.


\section{PCA of Ear Shapes} \label{sec:pcaEars}
From the set of ear shapes $E$ described in Section~\ref{sec:originalEarShapeDataset}, we classically constructed a statistical shape model of the ear using PCA \citep{cootes_active_1995}.
Let there be
$\mathbf{X}_{{E}} =
{ \left( \mathbf {e}_1 \; \dots \; \mathbf {e}_{{n}} \right) }^\mathrm{t} 
\in \mathbb{R}^{{n} \times {3 {n_v}}}$ 
the data matrix,
$\bar{\mathbf{e}} = \frac{1}{{n}}  \displaystyle\sum_{i=1}^{{n}} \mathbf{e}_i$
the average ear shape
and
$\bar{\mathbf{X}}_{E} = { \left( \bar{\mathbf{e}} \; \dots \; \bar{\mathbf{e}} \right) }^\mathrm{t} \in \mathbb{R}^{{n} \times {3 {n_v}}}$ the matrix constituted of the average shape stacked ${n}$ times.
Finally, let $\mathbf{\Gamma}_{{E}} \in \mathbb{R}^{3 {n_v} \times 3 {n_v}}$ be the covariance matrix of $\mathbf{X}_{{E}}$:
\begin{equation}
	\mathbf{\Gamma}_{{E}} = 
	\frac{1}{n - 1} 
	{\left( \mathbf{X}_{{E}} - \bar{\mathbf{X}}_{E} \right)}^\mathrm{t} 
	\left( \mathbf{X}_{{E}} - \bar{\mathbf{X}}_{E} \right) 
	.
\end{equation}

PCA can thus be written as
\begin{equation} 
	\label{eq:earPca1}
	\mathbf{Y}_{{E}} = \left( \mathbf{X}_{{E}} - \bar{\mathbf{X}}_{E} \right) {\mathbf{U}_{{E}}}^\mathrm{t} ,
\end{equation}
where $\mathbf{U}_{{E}}$ is obtained by diagonalizing the covariance matrix $\mathbf{\Gamma}_{{E}}$
\begin{equation}
	\label{eq:earPca2}
	\mathbf{\Gamma}_{{E}} = 
	\mathbf{U}_{{E}}^\mathrm{t} 
	{\mathbf{\Sigma}_{{E}}}^2 
	{\mathbf{U}_{{E}}}   
	 .
\end{equation}
In the equations above, ${\mathbf{\Sigma}_{{E}}}^2 \in \mathbb{R}^{(n-1) \times (n-1)}$ is the diagonal matrix that contains the eigenvalues of $\mathbf{\Gamma}_{{E}}$, ${\sigma_{{E}_1}}^2 , \, {\sigma_{{E}_2}}^2 , \, \dots {\sigma_{{E}_{{n} - 1}}}^2$, ordered so that ${\sigma_{{E}_1}}^2 \geq {\sigma_{{E}_2}}^2 \geq \dots \geq {\sigma_{{E}_{{n} - 1}}}^2$
\begin{equation}
	{\mathbf{\Sigma}_{{E}}}^2 = 
	\begin{bmatrix}
		{\sigma_{{E}_1}}^2 & & \\
		 & \ddots & \\
		& & {\sigma_{{E}_{{n} - 1}}}^2 \\
	\end{bmatrix}
	,
\end{equation}
and
$\mathbf{U}_{{E}} \in \mathbb{R}^{({n} - 1) \times 3 {n_v}}$ is an orthogonal matrix that contains the corresponding eigenvectors ${\mathbf{u}_{{E}_1}} , \, {\mathbf{u}_{{E}_2}}, \; \dots \; {\mathbf{u}_{{E}_{{n}-1}}} \in \mathbb{R}^{3 {n_v}}$ 
\begin{equation}
\label{eq:earPca3}
	\mathbf{U}_{{E}} = 
	\begin{bmatrix}
		{\mathbf{u}_{{E}_1}} \\
		\vdots \\
		{\mathbf{u}_{{E}_{{n}-1}}}
	\end{bmatrix}
	. 
\end{equation}
The eigenvalues denote how much variance in the input data is explained by the corresponding eigenvectors.

In the equations above, we implicitly set the number of principal components (PCs) to ${n}-1$, because all PCs after the $({n}-1)^{th}$ are trivial, i.e. of null associated eigenvalue.
Indeed, the number of examples $n$ is lower than the data dimension $3 n_v$ and the data is centered, thus
\begin{equation}
	r = \mathrm{rank} \left(  \mathbf{X}_{{E}} - \bar{\mathbf{X}}_{E} \right) \leq {n}-1
	.
\end{equation}
Hence, the rank of the covariance matrix does not exceed ${n}-1$ either:
\begin{equation}
	\mathrm{rank} \left( \mathbf{\Gamma}_E \right) \leq \min \left( r, r \right) = r \leq {n}-1
	.
\end{equation}

\begin{figure*} 
	\includegraphics[width=1.0\textwidth, trim = {2.3cm 12mm 6mm 0}]{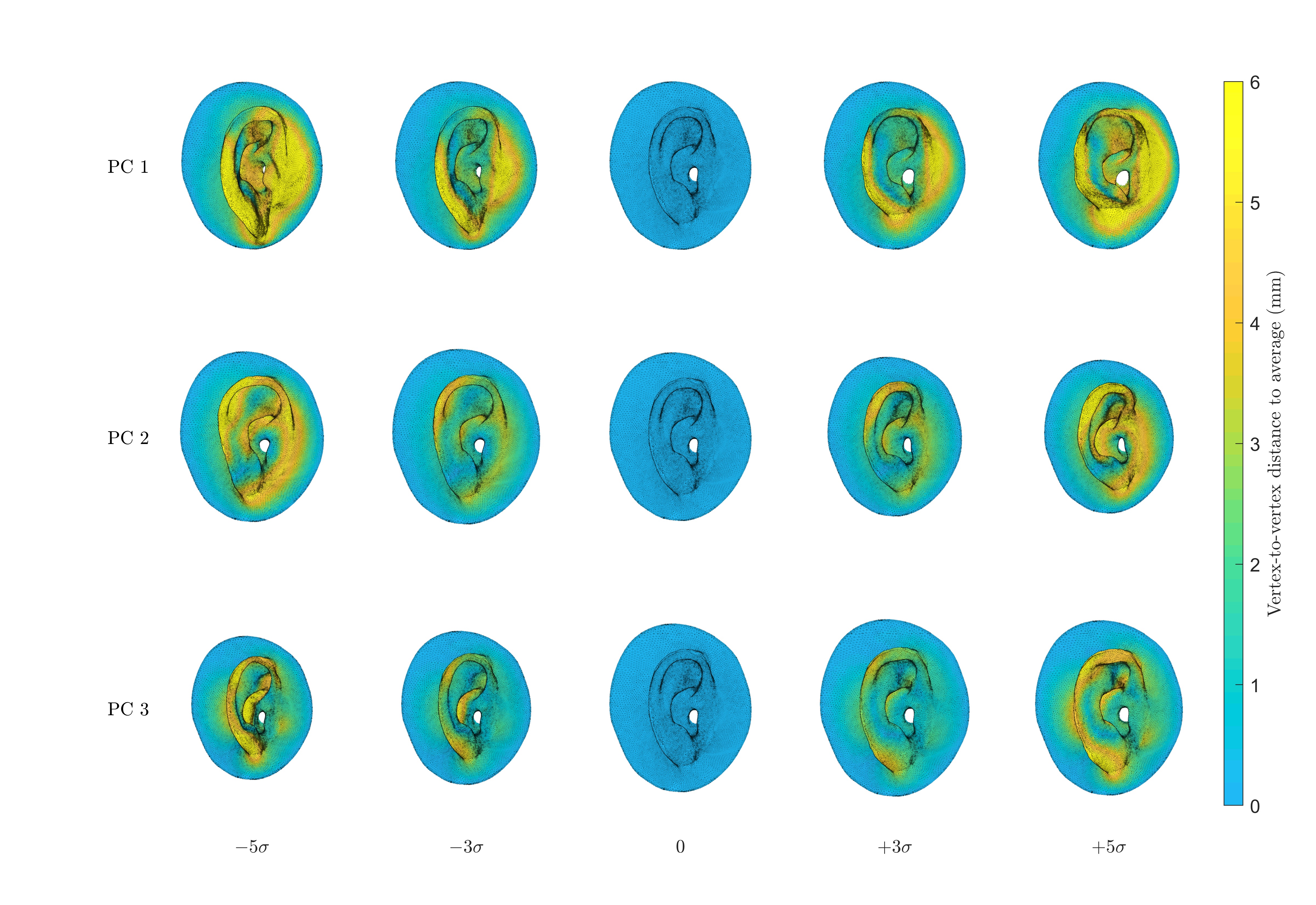}
	\caption{(color online) First three principal components (PCs) of the PCA ear shape model. Rows: PC of index $j \in \lbrace 1, \, 2, \, 3 \rbrace$. Columns: Weight assigned to given PC, indicated in proportion of its standard deviation $\sigma_{E_j}$.}
	\label{fig:3}
\end{figure*}

The behavior of the first 3 principal components is illustrated as follows.
For each PC of index $j \in \lbrace 1, \, 2, \, 3 \rbrace$, we set the $j^\mathrm{th}$ PC weight to $\lambda \sigma_{E_j}$ and all other PC weights to zero, with $\lambda \in \lbrace -5, \, -3, \, -1, \, +1, \, +3, \, +5 \rbrace$ and reconstructed the corresponding ear shape $\mathbf{e}_{v_j}  (\lambda)$ by inverting Equation~\eqref{eq:earPca1}
\begin{equation} \label{eq:earPcsBehavior}
	\mathbf{e}_{v_j}  (\lambda) = 
	\left( 0  \; \dots \; 0 \; {\lambda \sigma_{E_j}} \; 0 \; \dots \; 0 \right)
	\mathbf{U}_{E} + \bar{\mathbf{e}} 
	.
\end{equation}
Meshes derived from said ear shapes are displayed in Figure~\ref{fig:3}, colored with the vertex-to-vertex euclidean distance to the average shape.

The first one seems to control vertical pinna elongation including concha height and lobe length up to disappearance, as well as some pinna vertical axis rotation.
The second one seems to encode the intensity of some topography features such as triangular fossa depth or helix prominence. It also has an impact on concha shape and vertical axis rotation.
The third PC seems to have a strong influence on concha depth, triangular fossa depth as well as upper helix shape.

\section{PCA of Log-Magnitude PRTFs} \label{sec:pcaPrtfs}
In the following, we focus on the log-magnitude spectrum of the PRTFs. 
One reason is that HRTFs can be well modeled by a combination of minimum phase spectrum and pure delay \citep{kulkarni_minimum-phase_1995}.
Another one is the fact that, the TOA due to the pinnae, i.e. the one contained in PRTFs, is negligible compared to the effect of head and torso shadowing in HRTFs.
The logarithmic scale was chosen for its coherence with human perception.

For all ear shape $\mathbf{e}_i \in E$, let $\mathbf{p}_i = \varphi \left( \mathbf{e}_i \right) \in \mathbb{C}^{{n_f} \times {n_d}}$ be the corresponding PRTF set, computed according to the process described in Section~\ref{sec:numericalSimulations},
and let $\mathbf{q}_i = 20 \cdot \log_{10} \left( \lvert \mathbf{p}_i \rvert \right) \in \mathbb{R}^{{n_f} \times {n_d}}$ be the corresponding log-magnitude PRTF set, where the $\lvert \cdot \rvert$ and $\log_{10}$ operators are considered element-wise.
Accordingly, let
$\phi: \mathbb{R}^{3 {n_v}}  \longrightarrow  {\mathbb{R}}^{{n_f} \times {n_d}}$, defined by $\mathbf{e}  \longmapsto \mathbf{q} = 20 \cdot \log_{10} \left( \lvert \varphi(\mathbf{e}) \rvert \right)$, 
be the process of deriving a log-magnitude PRTF set from an ear point cloud, 
and let ${Q} = \left\lbrace \mathbf{q}_1, \; \dots \; \mathbf{q}_{{n}} \right\rbrace = \left\lbrace \phi(\mathbf{e}_1) \; \dots \; \phi(\mathbf{e}_{{n}}) \right\rbrace$ be the set of log-magnitude PRTF sets derived from $E$.


Most work in the literature either stacks the HRTFs of various directions and subjects into a data matrix of size $({n} \cdot n_d) \times n_f$ prior to PCA \citep{kistler_model_1992, middlebrooks_observations_1992}, or performs PCA one direction at a time on $n_d$ different ${n} \times n_f$-sized matrices \citep{nishino_estimation_2007, xu_improved_2008}.
In contrast, we chose to concatenate PRTFs from the $n_d$ directions into a row vector $\mathbf{q}_i \in \mathbb{R}^{n_f n_d}$ for each subject $i = 1, \; \dots \; {n}$.
The ${n}$ row vectors were then stacked into the data matrix $\mathbf{X}_{{Q}} = {\left( \mathbf{q}_1, \; \dots \; \mathbf{q}_{{n}} \right)}^\mathrm{t} \in \mathbb{R}^{{n} \times ({n_f} {n_d})}$. 
This approach has the advantage of parsing only the across-subject variability, instead of mixing the contributions of directionality and inter-individuality into the statistical analysis.

As in the case of ear shapes,
we performed PCA on the data matrix $\mathbf{X}_{{Q}}$ according to Equations~\eqref{eq:earPca1}, \eqref{eq:earPca2} and \eqref{eq:earPca3}.
The number of non-trivial PCs is $({n}-1)$ in this case as well, due to the fact that ${n} < n_f n_d$.

\begin{figure*} 
	\includegraphics[width=1\textwidth, trim={7mm 6mm 0 0}]{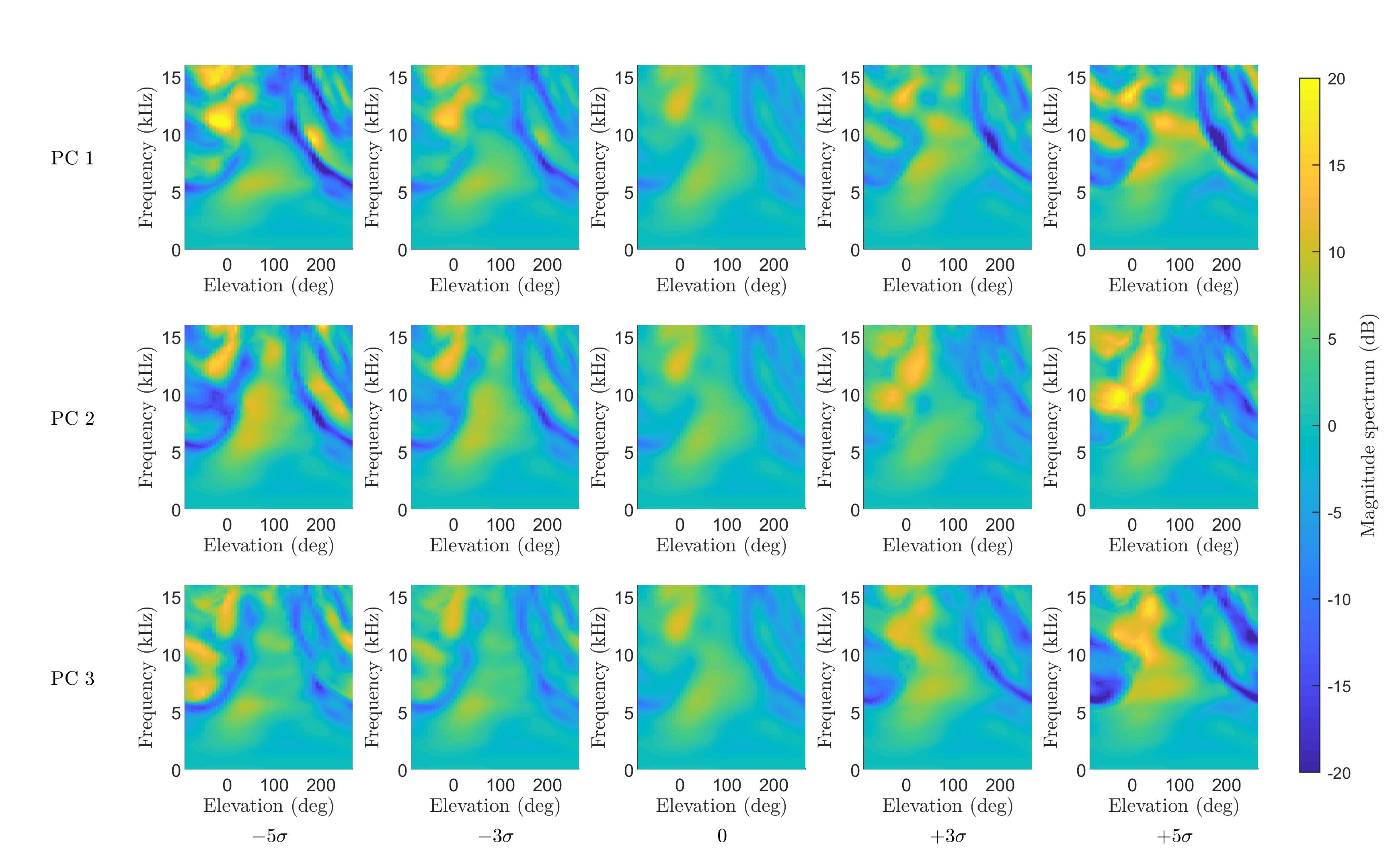} 
	\caption{
	(color online)
	First Principal Components (PCs) of the PCA model of log-magnitude PRTFs.
	Reconstructed PRTF sets are plotted in the median sagittal plane.
	Rows: PC. 
	Columns: Weight assigned to given PC, indicated in proportion of its standard deviation $\sigma$.}
	\label{fig:4}
\end{figure*}

Various PRTF sets that illustrate the behavior of the three first PCs were reconstructed as explained in Section~\ref{sec:pcaEars} and Equation~\eqref{eq:earPcsBehavior}.
They can be observed in Figure~\ref{fig:4} for directions that belong to the median sagittal plane. 
As it was expected, no variations are visible below 5~kHz: at these wavelengths the pinna have little impact on sound propagation.
Each PC appears to represent a different pattern of change in anterior and posterior directions, although only the first one seems to have a strong influence on directions above the head.
However, it does not seem possible to distinguish patterns that are limited to a certain range of directions and/or frequencies.
Furthermore, we are not able to identify a PC or a combination of PCs that represents a frequency shift in the PRTFs. 
Although the pinnae used to construct the model are normalized in size, one could have expected to observe frequency shifts due to variations in concha volume, for example.

\section{Comparison of Both PCA Models} \label{sec:compPcaModels}
\subsection{Dimensionality reduction capacity}
Let $S \in \left\lbrace E , {Q} \right\rbrace$ be either dataset. 
PCA can be used as a dimensionality reduction technique by retaining only the first $p$ PCs and setting the weights of the discarded PCs to zero \citep{jolliffe_principal_2002}, where $p \in \lbrace 1, \; \dots \; n-1 \rbrace$:
\begin{equation} \label{eq:pcaDimRed}
	 \tilde{\mathbf{Y}}_S = 
	\begin{bmatrix}
		y_{S_{1, 1}} & \dots   & y_{S_{1, p}} & 0          & \dots   & 0 \\
		\vdots  & \ddots & \vdots  &  \vdots & \ddots & \vdots \\
		y_{S_{n, 1}} & \dots   &y_{S_{n, p}} & 0           & \dots   & 0 \\
	\end{bmatrix}
	,
\end{equation}
where $y_{S_{i, j}}$ is the value of matrix $\mathbf{Y}_S$ at the $i^\mathrm{th}$ row and $j^\mathrm{th}$ column for all $i = 1, \; \dots \; {n}$ and $j = 1, \; \dots \; {{n} - 1}$.

Indeed, the change of basis defined by ${\mathbf{U}_S}^\mathrm{t}$ allows us to transform the dataset $\mathbf{X}_S$ into a domain where the associated covariance matrix ${\mathbf{\Sigma}_S}^2$ is diagonal with its diagonal values in decreasing order.
In other words, PCs are independent up to the second-order statistical moment and are ordered so that the first PCs describe more variability in the data than the last ones.

Approximated data can then be reconstructed by inverting Equation~\eqref{eq:earPca1}:
\begin{equation}
	\tilde{\mathbf{X}}_S = \tilde{\mathbf{Y}}_S {\mathbf{U}_{\mathrm{S}}} + \bar{\mathbf{X}}_S
	.
\end{equation}

\begin{figure} 
	\includegraphics[width=0.48\textwidth]{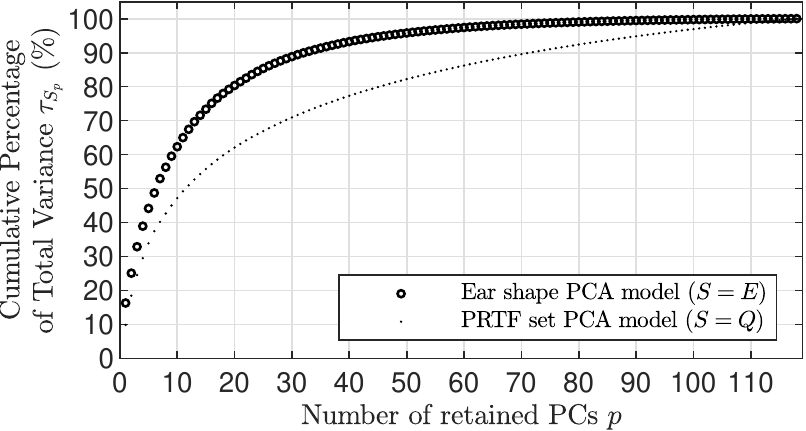}
	\caption{CPV $\tau_{S_p}$ as a function of the number of retained PCs $p \in \lbrace 1, \; \dots \;  {n} - 1 \rbrace$ for either PCA model. 
	                Circles: ear shape model ($S = E$). 
	                Dots: PRTF set model ($S = Q$).}
	\label{fig:5}
\end{figure}

A simple but useful metric to evaluate the capacity of a PCA model to reduce dimensionality is the cumulative percentages of total variance (CPV)  \citep[section 6.1]{jolliffe_principal_2002}
\begin{equation}
{\tau_S}_p = 
100 \cdot 
\left( \displaystyle \sum_{j=1}^{p} {{\sigma_S}_j}^2 \right)  
/ 
\left( {\displaystyle\sum_{j=1}^{{n} - 1} {{\sigma_S}_j}^2} \right)
, \;
\end{equation}
where $S \in \left\lbrace E, {Q} \right\rbrace$ represents either the set of ear shapes $E$ or the set of log-magnitude PRTFs ${Q}$ 
and $p \in \lbrace 1 , \; \dots \; {n} - 1 \rbrace$ is the number of retained PCs.
CPVs for both models are plotted in Figure~\ref{fig:5}.

A first notable result is that, for the ear shape model, the $99 \%$-of-total-variance threshold is reached for $p = 80$ retained PCs, i.e. only $\frac{p}{n - 1} = \frac{80}{118} =  67.8 \%$ of the maximum number of PCs.
In other words, the $118$-dimensional linear subspace of $\mathbb{R}^{3 {n_v}} = \mathbb{R}^{56661}$ defined by  the $n = 119$ pinnae of our database can be described using only 80 parameters with reasonable reconstruction accuracy, in the sense of a vertex-to-vertex mean-square error.

More importantly, PCA appears to be significantly more successful at reducing the dimension of ear shapes $\mathbf{e}_i$ than that of PRTF sets generated from the same ear shapes $\mathbf{q}_i = \phi(\mathbf{e}_i)$. 
Indeed, the PRTF CPV is significantly lower than the ear shape CPV for any number of retained PCs.
For instance, the $99 \%$-of-total-variance threshold is reached for 112 PCs out of 118 for the PRTF model against 80 out of 118 for the ear shape one.

\subsection{Statistical distribution} \label{subsec:statisticalDistribution}
Furthermore, in order to get a better idea of the repartition of the data in both 118-dimensional linear subspaces, we tested the PCs of each model for multivariate normal distribution using Royston's test \cite{royston_techniques_1983}. 
The test was performed on the columns of the PC weights matrix $\mathbf{Y}_S$, where $S \in \lbrace {E}, {Q} \rbrace$ denotes the dataset.

The outcome of the test is an associated p-value of  $3.7 \%$ in the case of ear PCs, and $0.0 \%$ in the case of PRTF PCs,
where the p-value refers to the null hypothesis that the distribution is not multivariate normal.
In other words, the ear model's PC weights can be considered to be multivariate-normally distributed with a significance level of $3.7 \%$, while its PRTF counterpart's fail the test for any significance level.

\subsection{Summary}
Overall, it appears that PCA performs better at modeling and reducing the dimensionality of ear shapes than of the corresponding log-magnitude PRTF sets.

Hence, linear techniques such as PCA seem ill-suited to reduce the dimensionality of PRTF sets. 
As, in addition, the ear shape model's PC weights follow a multivariate normal distribution, it appears to be more suitable than its PRTF counterpart for generalization and the generation of new data.

\section{Database Generation} \label{sec:databaseGeneration}
Non-linear machine learning methods may thus be more suited to model PRTFs  sets than linear ones.
However, such more complex techniques usually require larger amounts of data.
Nevertheless, as mentioned in the introduction, currently available databases of HRTFs feature about $10^2$ subjects in the best case, for a data dimension of about $10^6$, that is a proportion of $10^{-4}$ of the data's dimension. 
Hence, we propose a scalable method to construct a large database of synthetic PRTF sets by using the ear shapes space as a back door where to generate relevant artificial data.

\subsection{Random drawing of ear shapes}
The statistical ear shape model learned from dataset $E$ presented in Section~\ref{sec:pcaEars} can be used as a generative model.
Indeed, based on the results from Section~\ref{sec:compPcaModels}, we assume hereafter that the model's PCs (i.e. the columns of $\mathbf{Y}_{E}$) 
are mutually statistically independent and
follow normal probability laws of zero mean and $\sigma_{{E}_j}$ standard deviation $\mathcal{N}(0, \sigma_{{E}_j})$, where $j \in \lbrace 1, \; \dots \; n - 1 \rbrace$ represents the PC index.

An arbitrarily large number $N$ of ear shapes $\mathbf{e}'_1, \; \dots \; \mathbf{e}'_{N} \in \mathbb{R}^{3 {n_v}}$ could thus be generated as follows.
First, for all $i = 1, \; \dots \; N$,
a PC weights vector $\mathbf{y}_{{E}_i} = (y_{{E}_{i, 1}}, \; \dots \; y_{{E}_{i, {n} - 1}} ) \in \mathbb{R}^{{n} - 1}$
was obtained by drawing the $({n} - 1)$ PC weights $y_{{E}_{i, 1}}, \; \dots \; y_{{E}_{i, {n} - 1}}$ independently according to their respective probability laws $\mathcal{N}(0, \sigma_{{E}_1}), \, \dots \mathcal{N}(0, \sigma_{{E}_{{n}-1}})$.
Second, 
the corresponding ear shapes were reconstructed by inverting Equation~\eqref{eq:earPca1}
\begin{equation}
 	\mathbf{X}_{{E}}' = \mathbf{U}_{{E}} \mathbf{Y}_{{E}}' + \bar{\mathbf{X}}_{E}
	,
\end{equation}
where $\mathbf{Y}_{{E}}' \in \mathbb{R}^{N \times ({n} - 1)}$ is the matrix whose rows are the $N$ PC weights vectors
\begin{equation}
	\mathbf{Y}_{{E}}' = 
	\begin{bmatrix}
		 \mathbf{y}_{{E}_1}' \\
		\vdots \\
		\mathbf{y}_{{E}_{{N}}}' \\
	\end{bmatrix}
	=
	\begin{bmatrix}
		y_{{E}_{1, 1}}' & \dots & y_{{E}_{1, {n} - 1}}' \\
		\vdots & \ddots & \vdots \\
		y_{{E}_{N, 1}}' & \dots & y_{{E}_{N, {n} - 1}}'  \\
	\end{bmatrix}
	,
\end{equation} 
and $\mathbf{X}_{{E}}'  \in \mathbb{R}^{N \times {3 {n_v}}}$ is the data matrix whose rows are the $N$ ear shapes $\mathbf{e}_1', \; \dots \; \mathbf{e}_N' \in \mathbb{R}^{3 n_v}$
\begin{equation}
\mathbf{X}_{{E}}' = 	
\begin{bmatrix}
		\mathbf{e}_1' \\
		\vdots \\
		\mathbf{e}_N' \\
\end{bmatrix}
.
\end{equation}

\subsection{Ear shapes quality check}
At the end of the ear shape generation process, meshes were derived from the point clouds as in the case of the original dataset (see Section~\ref{sec:originalEarShapeDataset}).
We then verified that the meshes were not aberrant and that they were fit for numerical simulation: 
any mesh that presented at least one self-intersecting face was left out. 

In total, $24 \%$ (320 out of 1325) of the meshes were discarded.
Performing the Royston's multivariate normality test on the 1325 randomly drawn ear PC weights then on the 1005 remaining ones, we observed a decrease in the significance level of the test from $4.8 \%$ to $0.8 \%$:  the it appears that the statistical distribution of the ear PC weights was degraded by the selection process.
However, when looking into the distribution of each PC of the selected ear shapes separately (using the Shapiro-Wilk univariate normality test with a significance level of $5 \%$), we observe that the 9 rejected PCs account only for $3.7 \%$ of the total variance. 

For simplicity, we consider further on that $N$ is the number of retained meshes i.e. ${N} = 1005$.

\subsection{Numerical simulation}
Finally, PRTF sets were numerically simulated from the ear shapes of the new set $E'$ according to the process described in Section~\ref{sec:numericalSimulations}
\begin{equation}
	\mathbf{p}'_j = \varphi \left( \mathbf{e}'_j \right), \quad \forall j = 1, \; \dots \; {N}.
\end{equation}
Computing time for the simulation of the 1005 PRTF sets was of 40 days on a workstation that features 12 CPU and 32 GB of RAM.

\subsection{Data visualization}
\begin{figure*} 
    \subfloat[]{
    	\includegraphics[width=0.98\textwidth]{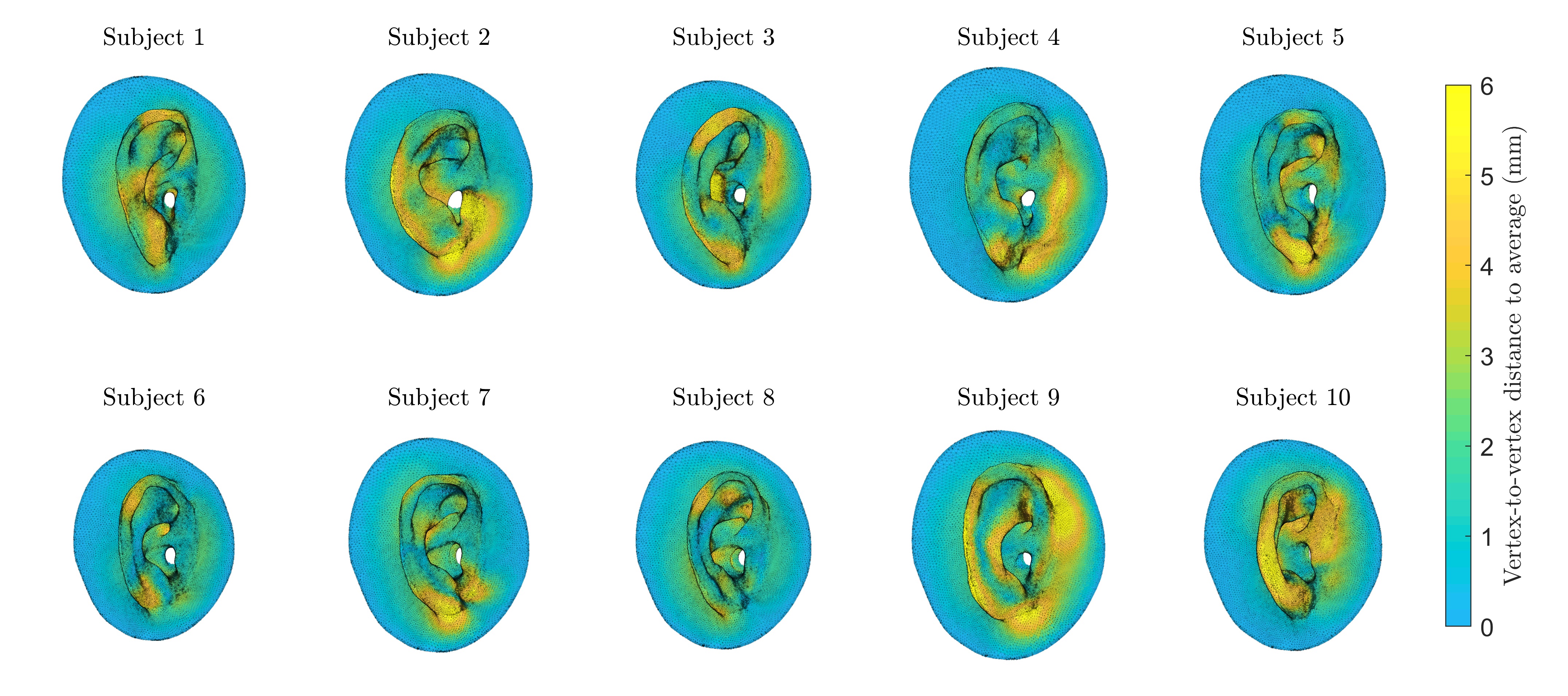}
    	}
	
    \subfloat[]{
    	\includegraphics[width=0.98\textwidth]{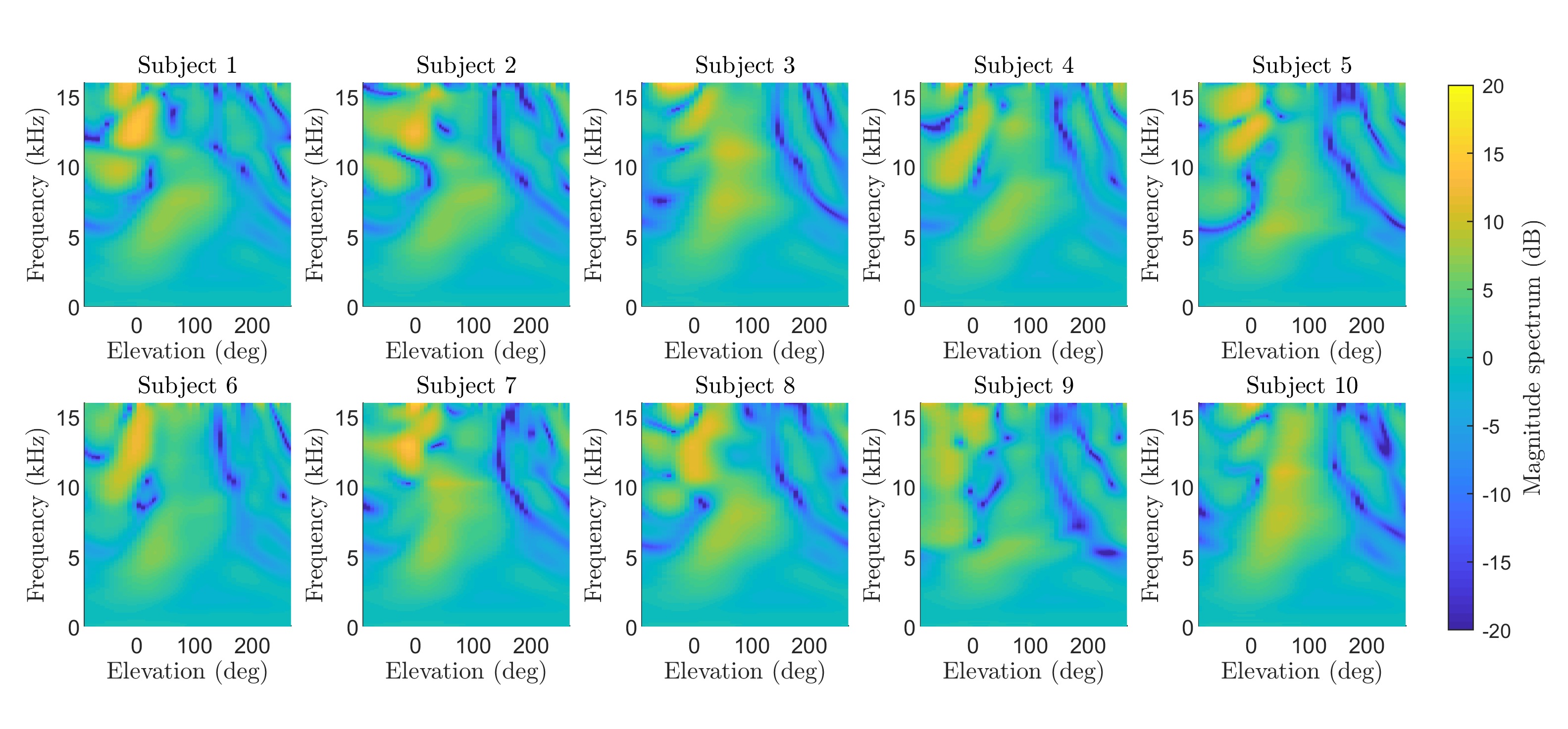}
    	}
\caption{
(color online)
Visualization of the first 10 subjects of WiDESPREaD.
(a) Meshes derived from the synthetic ear shapes $\mathbf{e}_1', \; \dots \; \mathbf{e}_{10}'$. Color represents the vertex-to-vertex euclidean distance to the generative model's average $\bar{\mathbf{e}}$. 
(b) Log-magnitude PRTF sets $20 \cdot \log_{10} ( \mathbf{p}_1' ), \; \dots \; 20 \cdot \log_{10} ( \mathbf{p}_{10}' )$ 
displayed in the median sagittal plane.
}
\label{fig:7}
\end{figure*}

By checking visually, we find that the synthesized ear shapes and PRTF sets look as realistic as hoped.
For purposes of illustration, the first 10 ear shapes and matching PRTF sets of the WiDESPREaD dataset are displayed in Figure~\ref{fig:7}.
These first 10 randomly drawn subjects illustrate well how ear shapes and matching PRTFs can be diverse and highlight the interest of this dataset.
 
\section{Conclusions}
In this paper, based on a proprietary dataset of 119 left ear meshes, we constituted a corresponding dataset of 119 PRTF sets by FM-BEM calculations.
We then applied a simple linear machine learning technique, PCA, independently to each dataset and found that it performed better at modeling and reducing the dimensionality of data on ear shapes than on PRTF sets.
Based on this result, we proposed a method to generate an arbitrarily large synthetic PRTF database by means of random drawing of ear shapes and FM-BEM calculation.
The resulting dataset of 1005 ear meshes and corresponding PRTF sets, named WiDESPREaD, is freely available on the Sofacoustics website: \url{https://www.sofacoustics.org/data/database/widespread}.

Increasing the number of PRTF sets by generating new artificial subjects in the ear shape space, where linear modeling seems adequate, may allow us to better understand the complexity of PRTF and HRTF generation from listener morphology and help to model them better. 
In particular, non-linear machine learning techniques such as neural networks can benefit from the scalability of this synthetic dataset generation process, as they usually require a large amount of data.
As it is, WiDESPREaD is the first database, to our knowledge, with over a thousand PRTF sets and matching registered ear meshes.
Although PRTFs are not complete HRTFs, they include an important part of the information relevant to HRTF individualization and,
as the dataset includes about 8 times more subjects than any available HRTF dataset, 
it has great potential to help develop and improve methods for HRTF modeling, dimensionality reduction and manifold learning, as well as spatial interpolation of sparsely measured HRTFs.

Future work includes the analysis of the augmented PRTF dataset and the search for a non-linear manifold.
If needed, new data can be generated to increase the size of the dataset, providing computing power and time. 
Furthermore, anthropometric measurements of the pinnae such as introduced with the CIPIC dataset \citep{algazi_cipic_2001} can be directly derived from the registered meshes, which may prove useful for the active field of HRTF individualization based on anthropometric features.
Finally, the method for data generation itself could be further improved on several aspects.
Indeed, our rudimentary generative ear shape model could be ameliorated by using either simple upgrades like probabilistic PCA \citep[p. 5]{luthi_statismo-framework_2012} or other modeling techniques altogether, although our results suggest that linear modeling techniques may be sufficient.
Going one step further, including statistical models of the human head, shoulders and right pinna could extend the method to the synthesis of HRTFs.

\section{Acknowledgments}
Special thanks go to our colleague Simon Leglaive for his thorough and insightful proofreading work.


\end{document}